\documentclass[journal]{IEEEtran}

\newlength{\figwidth} 
\usepackage[utf8x]{inputenc}
\usepackage[english]{babel}
\usepackage{verbatim} 
\usepackage{varioref}
\usepackage{psfrag}
\usepackage{epsf}
\usepackage{graphics}
\usepackage{graphicx}
\usepackage{amsbsy}
\usepackage{amssymb}
\usepackage{amscd}
\usepackage{amsmath}
\usepackage{amsthm}
\usepackage{enumerate} 
\usepackage{color}
\usepackage{epstopdf}
\usepackage{threeparttable}

\usepackage{cite}
\usepackage{url}

\newtheorem{theorem}{Theorem}

\newtheorem{remark}{Remark}

\title{Matching in the Air: Optimal Analog Beamforming under Angular Spread}

\author{ 
Jinfeng Du, \emph{Member, IEEE}, Marcin Rybakowski, Kamil Bechta, and Reinaldo A. Valenzuela, \emph{Fellow, IEEE} 
\thanks{Jinfeng Du and Reinaldo A. Valenzuela are with Nokia Bell Labs, Holmdel, NJ 07733, USA (e-mail: \{jinfeng.du, reinaldo.valenzuela\}@nokia-bell-labs.com).}
\thanks{Marcin Rybakowski and Kamil Bechta are with Nokia Mobile Networks, 54-130 Wroclaw, Poland (e-mail: \{marcin.rybakowski, kamil.bechta\}@nokia.com).}  
} 

\begin{document}

\maketitle

\begin{abstract}
Gbps wireless transmission over long distance at high frequency bands has great potential for 5G and beyond, as long as high beamforming gain could be delivered at affordable cost to combat the severe path loss. With limited number of RF chains, the effective beamwidth of a high gain antenna will be ``widened'' by channel angular spread, resulting in gain reduction. In this paper, we formulate the analog beamforming as a constrained optimization problem and present closed form solution that maximizes the effective beamforming gain. The optimal beam pattern of antenna array turns out to ``match'' the channel angular spread, and the effectiveness of the theoretical results has been verified by numerical evaluation via exhaustive search and system level simulation using 3D channel models. Furthermore, we propose an efficient angular spread estimation method using as few as three power measurements and validate its accuracy by lab measurements using a $16{\times}16$ phased array at 28 GHz. The capability of estimating angular spread and matching the beam pattern on the fly enables high effective gain using low cost analog/hybrid beamforming implementation, and we demonstrate a few examples where substantial gain can be achieved through array geometry optimization.
\end{abstract}
 


\section{Introduction}\label{sec:introduction}

	5G systems will adopt millimeter wave (mmWave) frequency bands to meet the capacity demand for future mobile broadband applications and new use cases\cite{TWC_01, TWC_02,TAP_4}. However, the high path loss and sensitivity to blockages~\cite{ TAP_2, TAP_15, TAP_3}, channel state information acquisition challenges~\cite{TAP_13}, hardware limitation and other difficulties~\cite{TAP_20}  make it challenging to provide high user rate at high frequencies without shrinking the traditional cell coverage range. 

\begin{figure} 
	\centering
		\includegraphics[width=0.9\figwidth]{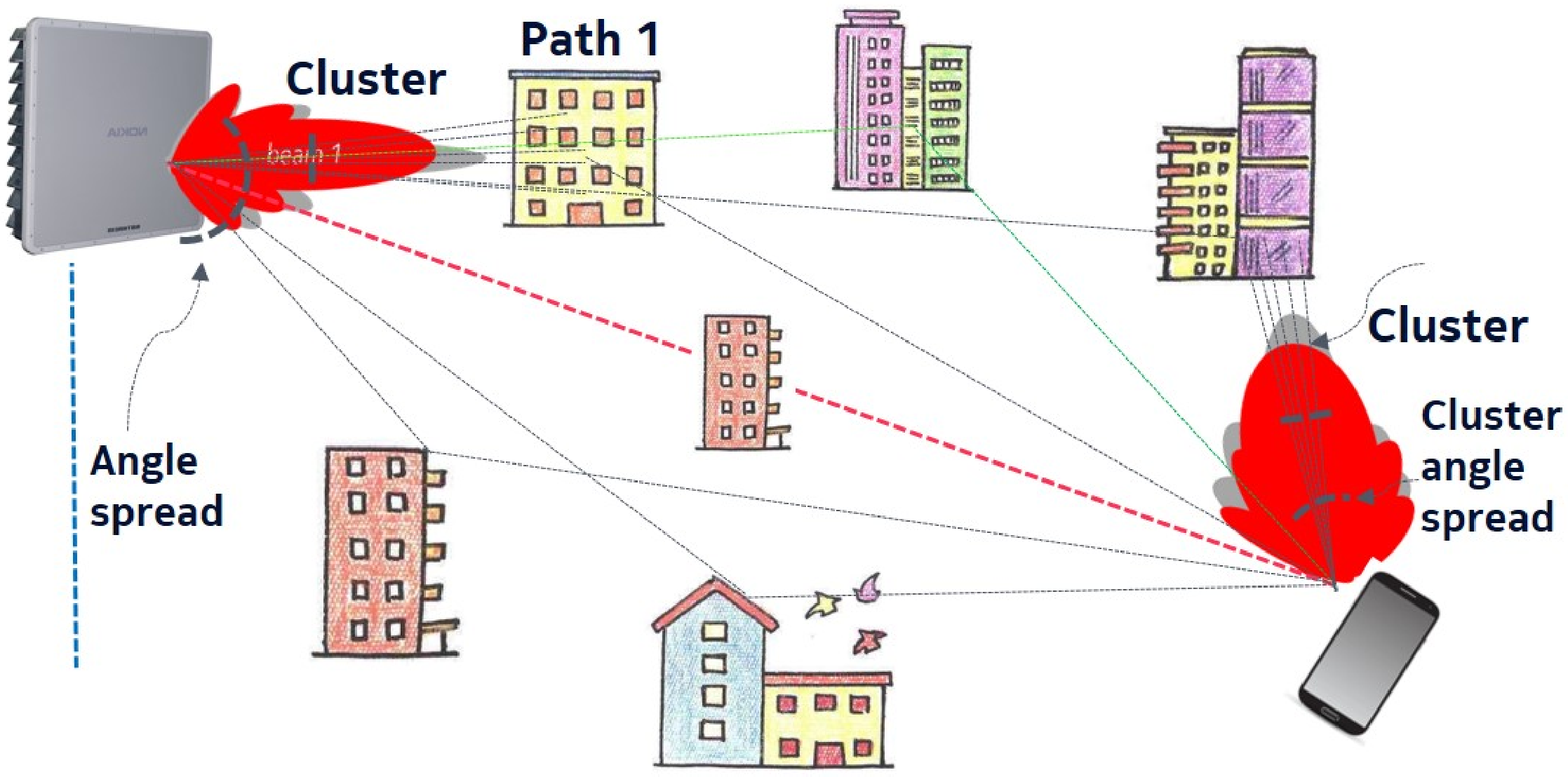}\\
		\includegraphics[width=0.65\figwidth]{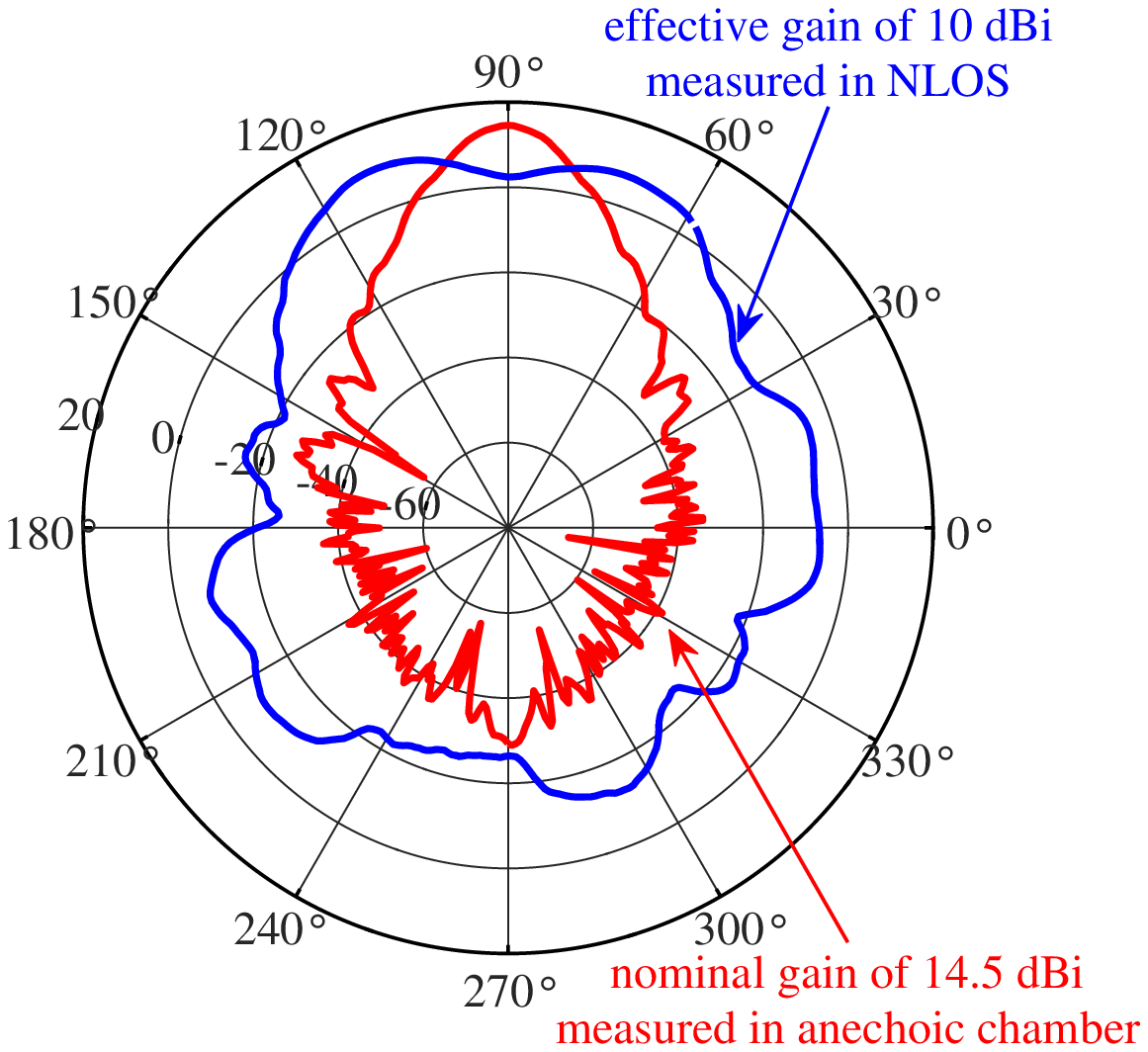}
	\caption{Illustration of angular spread in a NLOS multi-path propagation channel (upper) and the phenomenon of ``widened'' effective beamwidth from a NLOS measurement at 28 GHz (lower), where a 4.5 dB gain reduction has been observed compared to its nominal gain of 14.5 dBi.}
	\label{fig:effective_pattern}
\end{figure}

The critical part of high frequency links is the antenna and associated beamforming method. High beamforming gain is essential to combat the severe path loss such that Gbps throughput over long distance and coverage in non-line of sight (NLOS) areas can be realized. Full digital beamforming, capable of altering both amplitude and phase for each antenna element, is costly as it requires a dedicated RF chain for every antenna element and powerful baseband processing. Analog or hybrid beamforming with limited number of RF chains will be used in most of the products indented for mm-Wave frequency bands. However, owing to the channel angular spread and limited number of RF chains, the effective beamwidth of the antenna will be ``widened'' by the channel, as illustrated in Fig.~\ref{fig:effective_pattern}, resulting in reduced effective beamforming gain.  This can be intuitively understood by an analogy of lighthouse beacons being scattered in fog, leading to shortened reach. A sample measured beam pattern, presented in Fig.~\ref{fig:effective_pattern}, shows 4.5 dB gain reduction as compared to its nominal gain of 14.5 dBi (as measured in anechoic chamber). Previous measurement campaigns have reported significant loss of directional gain in various deployment scenarios, including suburban fixed wireless access (FWA)~\cite{TWC_2, TWC_4},  indoor offices~\cite{TWC_3}, and industrial factories~\cite{TWC_6}, where up to 7 dB gain reduction (90th percentile) out of 14.5 dBi nominal gain was reported.

\begin{figure} 
	\centering
		\includegraphics[width=0.9\figwidth]{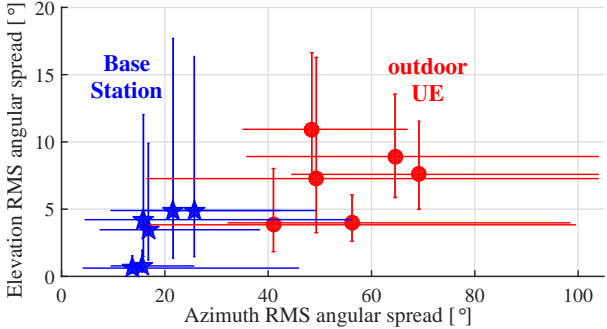}
	\caption{A chart of RMS angular spread (mean value and the corresponding range of 10\% to 90\%) for BS and for outdoor UE using 3GPP channel models~\cite{TWC_5} for 28 GHz with BS-UE distance of 100 m.}
	\label{fig:3GPP_spread}
\end{figure}

Angular spread has been widely acknowledged and carefully modeled for wireless communications, for example, by the 3rd Generation Partnership Project (3GPP)~\cite{TWC_5}. It is  different in azimuth and in elevation for most relevant deployment scenarios, and a chart of the root-mean-square (RMS) angular spread (its mean and associated 10\% to 90\% range) for base station (BS) and for outdoor user equipment (UE) is presented in Fig.~\ref{fig:3GPP_spread}, created based on 3GPP channel models~\cite{TWC_5} for 28 GHz with BS-UE distance of 100 m\footnote{Angular spreads are not sensitive to frequency or distance in~\cite{TWC_5}.}. Such difference has also been observed in other channel models developed by mmMagic, METIS, and NYU Wireless~\cite{R4-1706876}.

However, the impact of channel angular spread on system design, planning and performance evaluation has not been well understood. The prevailing practice for link budget calcualtion, inter-site interference and co-existence studies is to use nominal antenna pattern rather than the effective pattern, leading to inaccurate received power and interference level estimation.
Although high directional antennas have been used for backhaul links, they are usually installed at high heights with almost clear direct line-of-sight (LOS) path and close to zero angular spread. This is in contrast to mobile or fixed wireless access applications where the antennas might be below average clutter height 
and the impact of angular spread could be significant.  

\subsection{Our Contribution}

In this paper, we focus on wireless access deployment scenarios where large antenna arrays are deployed to improve the link budget. We take advantage of the difference in elevation and azimuth angular spread
and formulate the analog beamforming as a constrained optimization problem to maximize the effective beamforming gain.
We derive a closed form solution of the optimal array geometry, whose nominal beam pattern turns out to match the given channel angular spread. 
The potential gain of the optimal array over a square array of the same size is demonstrated by system level simulations using 3D channel models.
Furthermore, we also propose a method of estimating channel angular spread in azimuth and in evaluation using as few as three power measurements, and validate its accuracy via lab measurements using a $16{\times}16$ phased array at 28 GHz.   
The capability of estimating angular spread and optimizing beam pattern on the fly enables dynamic directional beam configuration, and it helps to achieve high effective gain using low cost analog/hybrid beamforming implementation. We also demonstrate a few examples where substantial gain can be achieved through array geometry optimization. To the best of our knowledge, this work is the first of its kind in matching antenna pattern with channel angular spreads to improve effective direction gain, which is essential and critical to ensure sufficient link budget in real deployment.

\subsection{Related Work}

 Some recent work have provided preliminary investigations on the impact of channel angular spread for channel modeling, link budget analysis, and system performance evaluation.
The mismatch between nominal antenna gain and received power level was observed in various channel measurements with directional antennas, and such antenna specific variation was embedded directly into ``directional'' path loss models~\cite{model_1, model_2, model_3}, which leads to different path loss models for each different combination of transmit and receive directive antennas. 
This is in contrast to the ``omni'' path loss models widely adopted by industrial standards such as~\cite{TWC_5} where 
the propagation channel is characterized free from any antenna assumptions and the path loss is modeled as it would be observed with ideal omni antennas at both the transmitter and the receiver. 
For example, in~\cite{TWC_3, TWC_4, TWC_6} the effective
gain reduction caused by angular spread is modeled separately from the ``omni'' path loss 
channel models. 
Reduction of directional gain and capacity by azimuth angular spread have been evaluated in \cite{JSAC13} for single/multiple sector beams, and the impact of angular spread in azimuth and in elevation for mmWave square arrays have been studied in \cite{TWC_1} for Gbps coverage with wireless relayed backhaul.

 System level simulations of mobile networks in \cite{TWC_04} have demonstrated up to 40\% deviation from realistic value of Long Term Evolution (LTE) downlink throughput when nominal antenna pattern is assumed instead of effective antenna pattern.
Study for 5G scenario with analog beamforming in mm-Wave range was presented in \cite{TWC_05} where the radio link budget for serving link and interfering links were evaluated for both nominal and effective antenna gains.
The impact of 3GPP 3D channel models on effective antenna array patterns has been visualized in \cite{TWC_06} and it was found that the downlink Signal to Interference and Noise  Ratio (SINR) can be overestimated by 10 to 17 dB in NLOS scenario when using nominal beam pattern rather than effective pattern.    
The impact of angular spread on the efficiency of tapering method has been evaluated via simulations \cite{TWC_07} which indicates that the first side-lobe suppression level  (SSL) can decrease to 16 dB in line of sight (LOS) conditions, or even to 2 dB in NLOS, in comparison to SSL of 20 dB for the nominal antenna pattern. 

\subsection{Paper Organization}

A brief description of system model is in Sec.~\ref{sec:model} and array geometry optimization  is presented in Sec.~\ref{sec:optimization}. System level simulation and lab measurements are reported in Sec.~\ref{sec:sim}. Several potential applications are discussed in Sec.~\ref{sec:application} and conclusions are in Sec.~\ref{sec:conclusion}. 

\section{System Models}\label{sec:model}   
 
To simplify presentation, we focus exclusively on beamforming over uniform planar array where elements are separated by half a wavelength. This configuration facilitates simple and direct representation of the nominal beam pattern by the underlining array size and array geometry. The same concept and methodology apply to other array types and beamforming methods where the RMS beamwidth of the beam pattern should be used for optimization.

We consider the case of high gain antennas whose beam pattern can be approximately characterized by Gaussian functions\footnote{Such approximation has been widely adopted in standard specifications such as 3GPP~\cite{TWC_5}. Empirical observation indicates that the main lobe can be well approximated by Gaussian function for antenna  gain as low as 5 dBi.}  both in azimuth and in elevation~\cite{TWC_1}
\begin{align}
g(\phi,\theta)=\frac{2}{B_h B_v}  e^{-\frac{\phi^2}{2B^2_h}}   e^{- \frac{\theta^2 }{2 B^2_v} },   \label{eqn:pattern}   
\end{align}
where $B_v$ and $B_h$ are the RMS beamwidth (in radius) in elevation and in azimuth, respectively.  
The directional gain, defined as the peak to average power ratio of the antenna pattern, is determined by the RMS beamwidths as~\cite{TWC_1} 
\begin{align}
G =\frac{2}{B_h B_v },   \label{eqn:3}                         
\end{align}

In the absence of scattering, the RMS beamwidths are set, correspondingly, to their nominal value $B_{v0}$ and $B_{h0}$, which can be determined from measurement in an anechoic chamber. 

 In the presence of scattering, signals may come from multiple directions. The received signal along a certain direction is the circular convolution of the nominal antenna pattern and the channel power angular response~\cite{TWC_3}. Assuming, for tractability, the channel angular spectrum of RMS azimuthal angular spread (ASD) $\sigma_h$ and RMS elevation angular spread (ZSD) $\sigma_v$ can be modeled as Gaussian functions with variance $\sigma_h^2$ and $\sigma_v^2$, respectively. The effective antenna pattern, which is a circular convolution of two independent Gaussian signals, still has the Gaussian form as \eqref{eqn:pattern} but with effective RMS beamwidth given by 
\begin{align}
B_v=\sqrt{B_{v0}^2+\sigma_v^2 },  \    B_h=\sqrt{B_{h0}^2+\sigma_h^2}.   \label{eqn:4}
\end{align} 
Therefore, we can determine the effective beamforming gain based on the nominal antenna pattern and channel angular spread. 
As a result, when the number of antenna elements increases, the effective gain in scattering environment is always smaller than its nominal gain, and will saturate\footnote{When there are as many RF chains as the number of antenna elements, generalized beamforming has the potential to provide effective gain that grows linearly with the number of elements, providing that perfect channel state information is available.} at the limit imposed by the channel angular spread.

 \section{Array Geometry Optimization and Angular Spread Estimation}\label{sec:optimization}
  
\subsection{Theoretical Derivation of Optimal Array Geometry}\label{sec:Geom_opt}

	We focus on analog/RF beamforming where there are in total $N$ antenna elements, arranged in rectangular/square shape to form a uniform planar array of size $(K_1, K_2)$, with 
	\begin{align}
	K_1K_2 \leq N. \label{eqn:5} 
\end{align}	
	Array of $(K_1, K_2){=}(1,N)$ corresponds to a horizontally deployed uniform linear array whereas $K_2{=}1$ indicates a vertically deployed uniform linear array. 
Since the effective beamforming gain depends on the panel geometry $(K_1, K_2)$, 
the nominal beamwidths $B_{ve}$ and  $B_{he}$ of the antenna elements, and channel angular spread $\sigma_h$ and $\sigma_v$,  we can optimize the array geometry $(K_1, K_2)$ to maximize the effective beamforming gain  subject to the size constraint \eqref{eqn:5}. 	
	 
\begin{theorem}
\label{theorem:gain}
Ignoring the integer constraint on array dimension $K_1$ and $K_2$, the effective beamforming gain of an antenna array with $N$ elements is upper bounded as
\begin{align} 
G(N, B_{ve},B_{he},\sigma_v,\sigma_h ) \leq \frac{2}{\sigma_h\sigma_v + \frac{B_{ve}B_{he} }{N}},
\end{align} 
with equality if and only if the array geometry is given by
\begin{align}
K_1=\sqrt{ \frac{N B_{ve}\sigma_h}{B_{he}\sigma_v }}, \   K_2=\sqrt{\frac{N B_{he}\sigma_v}{B_{ve}\sigma_h }}. \label{eqn:opt_geometry}
\end{align}
\end{theorem} 
\begin{IEEEproof}
See Appendix \ref{app:Proof}.
\end{IEEEproof}
 
The nearest integer pair close to $(K_1, K_2)$ as specified by \eqref{eqn:opt_geometry} and satisfying the total elements constraint \eqref{eqn:5} gives the best analog beamforming gain.

\begin{figure} 
	\centering
		\includegraphics[width=0.9\figwidth]{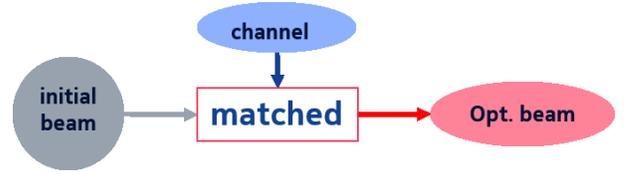}
	\caption{The optimal beam pattern (and the underlining array geometry using uniform plenary array) should match the channel angular spread as prescribed by \eqref{eqn:match} to maximize the effective analog beamforming gain. }
	\label{fig:match}
\end{figure}

Note that the ratio between the optimal RMS azimuth and elevation beamwidth equals the ratio of the channel RMS spread in azimuth and in elevation, i.e.,
\begin{align}
 \frac{B_{h0}}{B_{v0}} =\frac{B_{he}/K_2}{ B_{ve}/K_1} = \frac{\sigma_h}{\sigma_v}. \label{eqn:match}
\end{align} 
Hence, the optimal beam pattern (generated by the optimal array geometry) matches the channel angular spread in both azimuth and elevation, as illustrated in Fig.~\ref{fig:match}.

\begin{remark}\label{remark:implementation}
The optimal geometry that provides the maximal effective gain is determined for the given angular spread and number of elements. The actually implementation might not be exact as what suggested by the optimal solution due to implementation difficulties or cost constraints. For example, RF design would prefer symmetric circuits and antenna elements placement, and the use of splitters in the feed network may limit the granularity of array geometry options.  Nevertheless, the beam pattern should match the angular spread as close as possible as prescribed in \eqref{eqn:match} after balancing all the tradeoffs.  
\end{remark}

\begin{remark}\label{remark:other_antenna_types}
The array geometry optimization for uniform plenary arrays is also applicable for other types of directional antennas (like horn, reflector antennas, plasma antennas, etc.) or antenna arrays using non-directional elements (dipole, monopole, etc.) where the optimal antenna (array) is designed by optimizing the beam pattern in azimuth and in elevation to achieve the maximal effective antenna gain in a given channel. 
\end{remark}

\subsection{Theoretical Derivation of Angular Spread Estimation}\label{sec:spread_estimation}

	When channel angular spread  (ASD $\sigma_h$ and/or ZSD $\sigma_v$) is unknown or time varying, the effective gain of a rectangular-shaped sub-array can be determined in real time from measured signal strength using three or more different sub-array configurations, as detailed below. For a uniform planar array of size $(N_1, N_2)$, i.e., there are $N_1$ rows and $N_2$ column, we can measure the signal strength of three sub-panels of size $(n_1, k_1)$, $(n_1, k_2)$, and $(n_2, k_1)$, where $n_1, n_2\leq N_1$, and $k_1, k_2 \leq N_2$. The effective gains of the corresponding sub-arrays, which depend  on $(B_{ve},B_{he},\sigma_v,\sigma_h)$ but not shown explicitly to simplify notation, can be written as
	\begin{align}
G(n_1,k_1 )=\frac{2}{\sqrt{(B_{ve}/n_1)^2+\sigma_v^2 } \sqrt{(B_{he}/k_1)^2+\sigma_h^2 }}, \label{eqn:11} \\
G(n_1,k_2)=\frac{2}{\sqrt{(B_{ve}/n_1)^2+\sigma_v^2 } \sqrt{(B_{he}/k_2)^2+\sigma_h^2 }},     \label{eqn:12}\\
G(n_2,k_1)=\frac{2}{\sqrt{(B_{ve}/n_2)^2+\sigma_v^2 } \sqrt{(B_{he}/k_1)^2+\sigma_h^2 }}.     \label{eqn:13}
\end{align}
By combining \eqref{eqn:11} and \eqref{eqn:12} we have,
\begin{align}
 \frac{G(n_1,k_1)}{G(n_1,k_2)}
& =\frac{\sqrt{(B_{he}/k_2)^2+\sigma_h^2 }}{\sqrt{(B_{he}/k_1)^2+\sigma_h^2 }},  \label{eqn:14} 
\end{align}
from which we can obtain
\begin{align}
  \left[\frac{G^2(n_1,k_2)}{G^2(n_1,k_1)}-1\right]\left(\frac{\sigma_h}{B_{he}}\right)^2 = \frac{1}{k_1^2} -  \frac{G^2(n_1,k_2)}{k_2^2G^2(n_1,k_1)}, \label{eqn:15}
\end{align}
leading to an estimate of normalized ASD, in its squared form,   
\begin{align}
\left(\frac{\sigma_h}{B_{he}}\right)^2 = \frac{1/k_1^2 - {G^2(n_1,k_2)}/{(k_2^2G^2(n_1,k_1))}}{ {G^2(n_1,k_2)}/{G^2(n_1,k_1)}-1}. \label{eqn:16}
\end{align}

Similarly, by combining (11) and (13) we obtain an estimate of the normalized ZSD, in its squared form,as 
\begin{align}
  \left[\frac{G^2(n_2,k_1)}{G^2(n_1,k_1)}-1\right]\left(\frac{\sigma_v}{B_{ve}}\right)^2 = \frac{1}{n_1^2} -  \frac{G^2(n_2,k_1)}{n_2^2G^2(n_1,k_1)}, \label{eqn:17}\\
	\left(\frac{\sigma_v}{B_{ve}}\right)^2 = \frac{1/n_1^2 - {G^2(n_2,k_1)}/{(n_2^2G^2(n_1,k_1))}}{ {G^2(n_2,k_1)}/{G^2(n_1,k_1)}-1}. \label{eqn:18}
\end{align}

If there are more measurements using different sub-arrays, each such pair would provide an estimate of the normalized ASD or ZSD, and such estimates should be combined together by treating each of such estimation as one realization of \eqref{eqn:15} and \eqref{eqn:17} for ASD and ZSD, respectively. Then all the equations formulated using \eqref{eqn:15} will be treated as an overdetermined linear system for ASD and all the equations formulated using \eqref{eqn:17} will be treated as an overdetermined linear system for ZSD. Given $n$ independent measurements of ASD established by \eqref{eqn:15}, we donate $a_i$ and $b_i$ as the corresponding constant on the left-hand-side (LHS) and the right-hand-side (RHS), respectively, of ASD estimation \eqref{eqn:15}, for pair $i=1,\ldots,n$. Similarly, denote $c_j,b_j,j=1,\ldots,l$, as the LHS and RHS constants, respectively, of ZSD estimation \eqref{eqn:17}.
We will have
\begin{align}
 \left(\frac{\sigma_h}{B_{he}}\right)^2 \bar{\boldsymbol{a}}  = \bar{\boldsymbol{b}}, \ 
\left(\frac{\sigma_v}{B_{ve}}\right)^2 \bar{\boldsymbol{c}}  = \bar{\boldsymbol{d}},  \label{eqn:19}
\end{align}
where 
\begin{align*}
\bar{\boldsymbol{a}}\triangleq [a_1,\ldots,a_n]^T, \ \bar{\boldsymbol{b}}\triangleq [b_1,\ldots,b_n]^T, \\
\bar{\boldsymbol{c}}\triangleq [c_1,\ldots,c_l]^T, \ \bar{\boldsymbol{d}}\triangleq [d_1,\ldots,d_l]^T. 
\end{align*}
Then we can apply the classical Least Square estimator to obtain the improved estimation of the normalized ASD and ZSD, in their squared form, as
\begin{align}
\left(\frac{\sigma_h}{B_{he}}\right)^2 = \frac{\bar{\boldsymbol{a}}^T\bar{\boldsymbol{b}}}{\bar{\boldsymbol{a}}^T\bar{\boldsymbol{a}}}, \  
\left(\frac{\sigma_v}{B_{ve}}\right)^2 = \frac{\bar{\boldsymbol{c}}^T\bar{\boldsymbol{d}}}{\bar{\boldsymbol{c}}^T\bar{\boldsymbol{c}}}. \label{eqn:20}
\end{align}
Estimators other than the Lease Square estimator used here in \eqref{eqn:20} can also be applied here to tradeoff among accuracy, complexity and robustness.

Note that a legitimate estimate of the squared ASD and ZSD should always be non-negative, but the estimates obtained using \eqref{eqn:16}, \eqref{eqn:18}, or \eqref{eqn:20} might be negative because of estimation noise. Therefore, any of the estimates  whose value is negative should be replaced by zero. 

With estimation from \eqref{eqn:16}, \eqref{eqn:18}, or \eqref{eqn:20}, the effective gain of a sub-array of size $(m_1, m_2)$ can be estimated as
\begin{align}
{G(m_1,m_2)} = {G(n_1,k_1)}\frac{ \sqrt{\frac{1}{n_1^2} + \frac{\sigma_v^2}{B_{ve}^2}} \sqrt{\frac{1}{k_1^2} + \frac{\sigma_h^2}{B_{he}^2}}} {\sqrt{\frac{1}{m_1^2} + \frac{\sigma_v^2}{B_{ve}^2}} \sqrt{\frac{1}{m_2^2} + \frac{\sigma_h^2}{B_{he}^2}}}. \label{eqn:21}
\end{align}

\section{Numerical Evaluation, System Level Simulation and Lab Measurements}\label{sec:sim}
In this section we will demonstrate the benefits of array geometry optimization by numerical results, system level simulation and lab measurements using a 28 GHz phased array with 256 elements. 

\subsection{Numerical Evaluation}

\begin{figure} 
	\centering
		\includegraphics[width=0.9\figwidth]{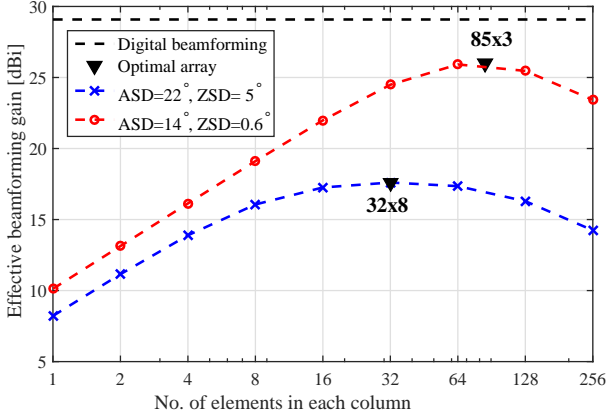}
	\caption{Effective beamforming gain of \eqref{eqn:effective_gain} for analog beamforming using uniform planar array with $256$   elements at 28 GHz with BS-UE distance of 100 meters for both the UMa NLOS scenario (blue line) and the UMi Street Canyon LOS scenario (red line) using 3GPP models~\cite{TWC_5}. The optimal array geometries from Theorem~\ref{theorem:gain} are highlighted as black triangles.}
	\label{fig:analog_BF}
\end{figure}

Effective beamforming gain for analog beamforming (i.e., one RF chain) using uniform planar arrays with $256$ antenna elements at 28 GHz  are shown in Fig.~\ref{fig:analog_BF} for both the UMa NLOS scenario (blue line) and the UMi Street Canyon LOS scenario (red line), where the angular spreads of radio channels are from 3GPP models~\cite{TWC_5} assuing BS-UE distance of 100 m. The effective gain obtained using \eqref{eqn:effective_gain} for a set of different array geometry are marked by markers and connected by solid curves  to illustrate the general trend of effective gain with respect to array geometry. The optimal array geometries for each channel, designed based on Theorem~\ref{theorem:gain}, are highlighted in the plot using black triangles. 
   
With total of 256 elements, 5dBi each, the ideal gain obtained by digital beamforming with full channel state information would be $29.1$ dBi. 
In scenarios where angular spread is moderate, such as the 3GPP UMi Street Canyon LOS with medium ASD of $14^\circ$ and ZSD $0.6^\circ$,  a $64\times 4$ tall array (very close to the optimal geometry of $85\times3$) is 4 dB better than the $16\times16$ square array, and 16 dB better than a $1\times 256$ fat array. In a different environment such as the 3GPP UMa NLOS case which is characterized by larger angular spreads (medium ASD of $22^\circ$ and ZSD $5^\circ$), a $32\times8$ tall array (optimal) is 0.5 dB better than a $16\times16$ square array, 9 dB better than a $1\times256$ fat array.  This shows how important it is to have matched antenna beam pattern to radio channels and highlights the benefit of adapting antenna beam pattern to particular angular spreads of radio channels.

\subsection{System Level Simulation Using 3D Channel Models}
The system level simulation was performed to examine the accuracy of the theoretical analysis presented in Sec.~\ref{sec:optimization} with full 3D spatial statistical channel model, as specified in 3GPP TR 38.901~\cite{TWC_5}, and antenna array model with beamforming algorithm adopted from 3GPP 5G system evaluation described in 3GPP TR 38.803~\cite{TWC_5b}. Key parameters of our system level simulation are summarized in Table~\ref{tab:sim_setup}.

\begin{table}[t]
	\centering 
	\caption{Summary of key parameters of the system level simulation.}
		\begin{tabular}{|c|l|}
		\hline
Parameters 	& Values\\ 
	\hline
Network layout & 3-ring hexagon-grid with wrap around, 200 m ISD \\
\hline
Macro cell & 19 sites, each has 3 ``cell'' (location anchor, no BS)\\ 
\hline
Micro BS & 3 cluster circles per macro; each has 1 micro BS\\
\hline
BS drop & Random drop along the edge of cluster circles\\
\hline			
BS antenna & Uniform planar array with 128 elements (8 dBi)\\
\hline
Antenna pattern & as per 3GPP TR 38.803~\cite{TWC_5b}\\
\hline
BS antenna height & 10 m\\
\hline
UE height & 1.5 to 22.5 m\\
\hline
Number of UE &  1 per micro BS\\
\hline
UE location & 20\% outdoor, 80\% indoor\\
\hline
Penetration loss & 50\% high loss, 50\% low loss\\
\hline
UE distribution & uniform\\
\hline 
BS-UE distance & minimum 3 m (2D)\\
\hline
LOS probability & as per 3GPP TR 38.901~\cite{TWC_5}\\
\hline
Channel model & 3GPP TR 38.901 UMi Street Canyon\\
\hline
Correlation &  0.5 between sites\\
\hline
		\end{tabular}
	\label{tab:sim_setup}
\end{table}

\begin{table}[t]
	\centering	
	\caption{Simulation results match theoretical analysis of effective beamforming gain of rectangular arrays with maximum of 128 elements (each of 8 dBi gain).}
	\label{tab:simulation}
		\begin{tabular}{|c|c|c|c|}
		\hline
BF gain	[dBi] &  Nominal &	Analysis 	& Simulation\\ 
	\hline
 $8	\times 16$ 	& 29.07	& 19.91 &	19.58\\
\hline
$42	\times 3$	& 29.00	& 24.31	& 24.35\\
\hline			
		\end{tabular}
\end{table}

	\begin{figure}  
	\centering
		\includegraphics[width=0.9\figwidth]{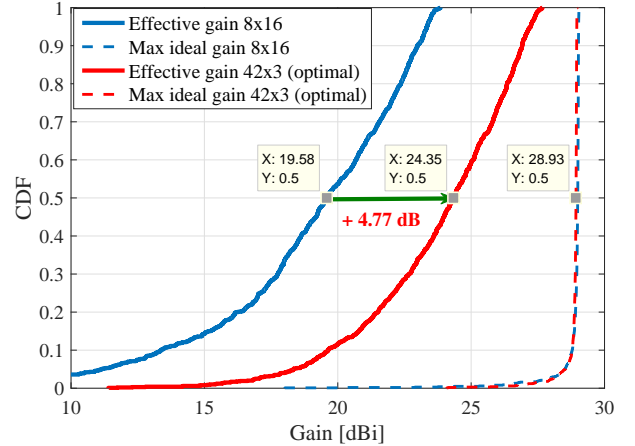}
	\caption{System level simulation results of the effective beamforming gain of rectangular arrays with maximum 128 elements using 3GPP 3D spatial channel model with medium ASD of $16^\circ$ and  ZSD of $1^\circ$ for both the default array geometry of $8\times16$ (blue lines) and the optimal geometry of $42\times3$ (red lines). The median of the gains obtained from system level simulation match the values predicted by theoretical analysis within 0.5 dB.}
	\label{fig:analog_BF_sim}
\end{figure}

First set of simulation results aim to verify correctness of analysis of effective antenna gain, for BS transmission in downlink, described above. For this purpose, we override some of the simulation parameters from Table~\ref{tab:sim_setup} to remove some constraints normally seen in system level simulations. More specifically, we set all UEs at 10 m high (same height as the BS) and 60 m from its serving BS with both the BS and UE antennas aiming towards the strongest direction on its boresight. The means ASD is fixed to $16^\circ$ and mean ZSD of $1^\circ$ to facilitate direct comparison against theoretical analysis.   Results of simulations are presented in Fig.~\ref{fig:analog_BF_sim} and Table~\ref{tab:simulation}. It can be noticed, that median value of antenna gain CDF matches the analytical effective antenna gain within 0.5 dB.

\begin{figure*} 
	\centering
		\includegraphics[width=0.9\textwidth]{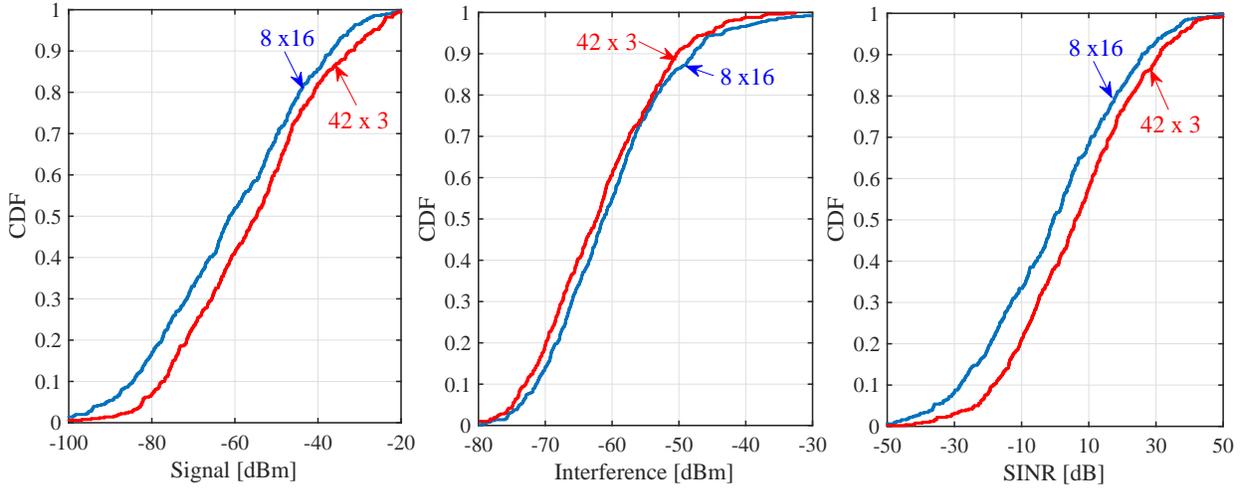}
	\caption{System level simulation results of the DL received signal power (left), interference power (middle), and the SINR (right) of rectangular arrays for both the default array geometry of $8\times16$ (blue lines) and the optimized geometry of $42\times3$ (red lines). 3GPP 3D spatial channel model under UMi Street Canyon scenario. The combined gain of signal power increase and interference power decrease leads to an increase of median SINR by 6.6 dB.}
	\label{fig:analog_BF_sim_UMi}
\end{figure*}

Second set of simulation results are to demonstrate the benefits of optimizing antenna array geometry in realistic deployment scenarios as described in Table~\ref{tab:sim_setup}.  Two array geometries are used in simulation, i.e., the default $8{\times}16$ arrangement and the optimal $42{\times}3$ configuration as obtained using Theorem~\ref{theorem:gain}. Simulation results for the received DL serving signal power, DL interference power, and DL SINR are presented in Fig.~\ref{fig:analog_BF_sim_UMi}. As compared to the default  $8{\times}16$  array configuration assumed by 3GPP,  the optimized  $42\times3$ array has demonstrated large increase in signal power (Fig.~\ref{fig:analog_BF_sim_UMi} left) thanks to its matching to channel angular spread, and modest reduction in interference power (Fig.~\ref{fig:analog_BF_sim_UMi} middle) thanks to its increased vertical resolution, leading to a combined gain of 6.6 dB on median SINR (Fig.~\ref{fig:analog_BF_sim_UMi} right).
Should all users/devices distributed at the same height, widened azimuthal beam may lead to an increase in interference and therefore smaller SINR gain using optimized array geometry.  

\subsection{Lab Measurements}

	\begin{figure} 
	\centering
		\includegraphics[width=0.34\figwidth]{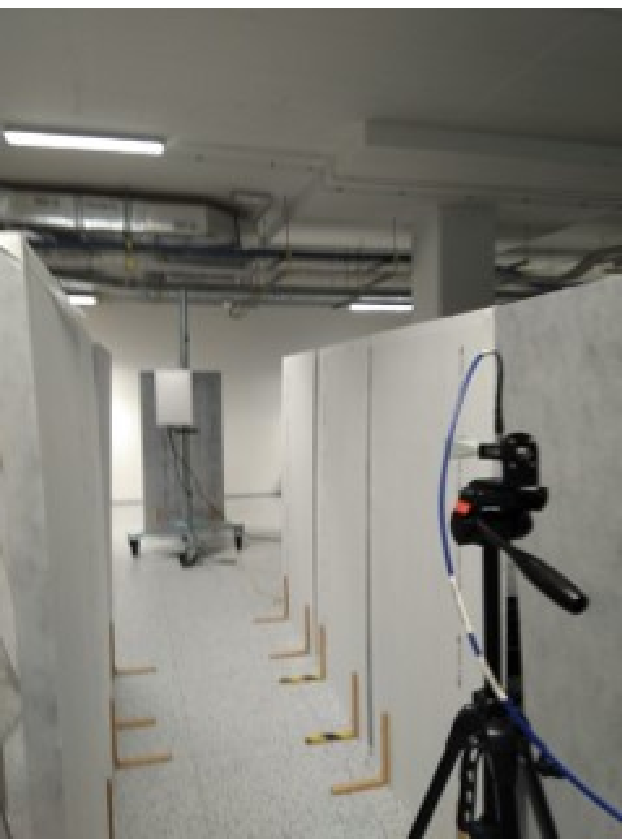} \includegraphics[width=0.61\figwidth]{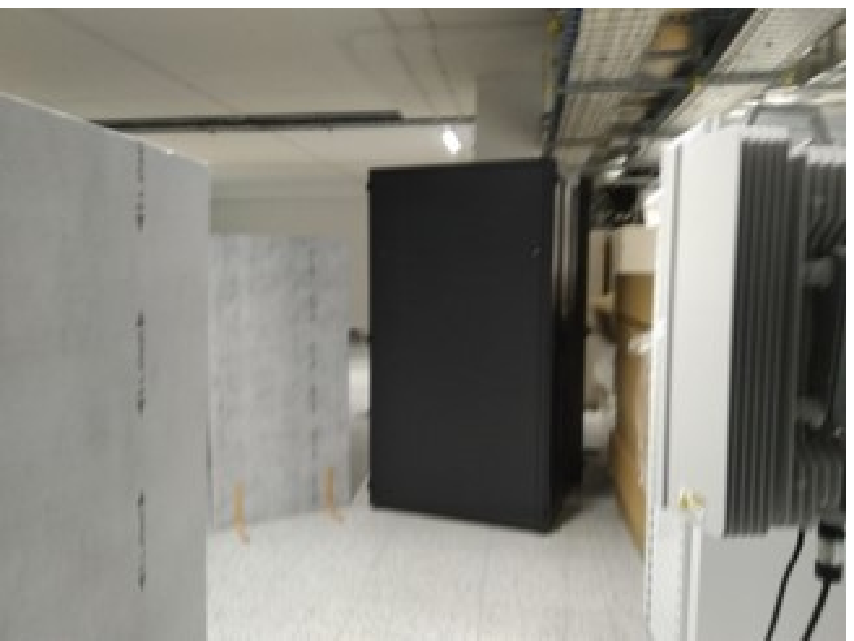}
	\caption{Lab measurement setup for both LOS (left) and NLOS (right).}
	\label{fig:lab_meas_setup}
\end{figure}

Lab measurements were carried out using a 28 GHz $16{\times}16$ array as the transmitter (Tx) and a 10 dBi horn as the receiver (Rx). Different antenna array geometry was configured by setting zero amplitude for selected antenna elements (AE). The ``muted'' antenna elements behaved like dummy elements which have marginal impact for antenna pattern due to EM coupling from active AEs. However, this small impact does not influence our general conclusion. The Rx horn antenna was connected to a Signal Analyzer. Tx signal with 100~MHz bandwidth was radiated from the antenna array and the received signal power was measured at Rx side. Since different Tx sub-array has different Tx power, the difference in beamforming gain is determined by the difference in Rx power subtracting the difference in Tx power. This operation also eliminates the common losses (such as cable loss, connector loss) experienced by all signals. 

Calibration in anechoic chamber was done using different antenna array configurations with boresight alignment. The measured total array gain with the same number of antenna elements but different geometry (e.g. $8\times8$, $16\times4$, $4\times16$ for 64  elements) was almost the same, with difference around 0.5 dB which could be attributed to dummy elements coupling effect, beam alignment offset or other measurement noise.

\begin{figure} 
	\centering
		\includegraphics[width=0.95\figwidth]{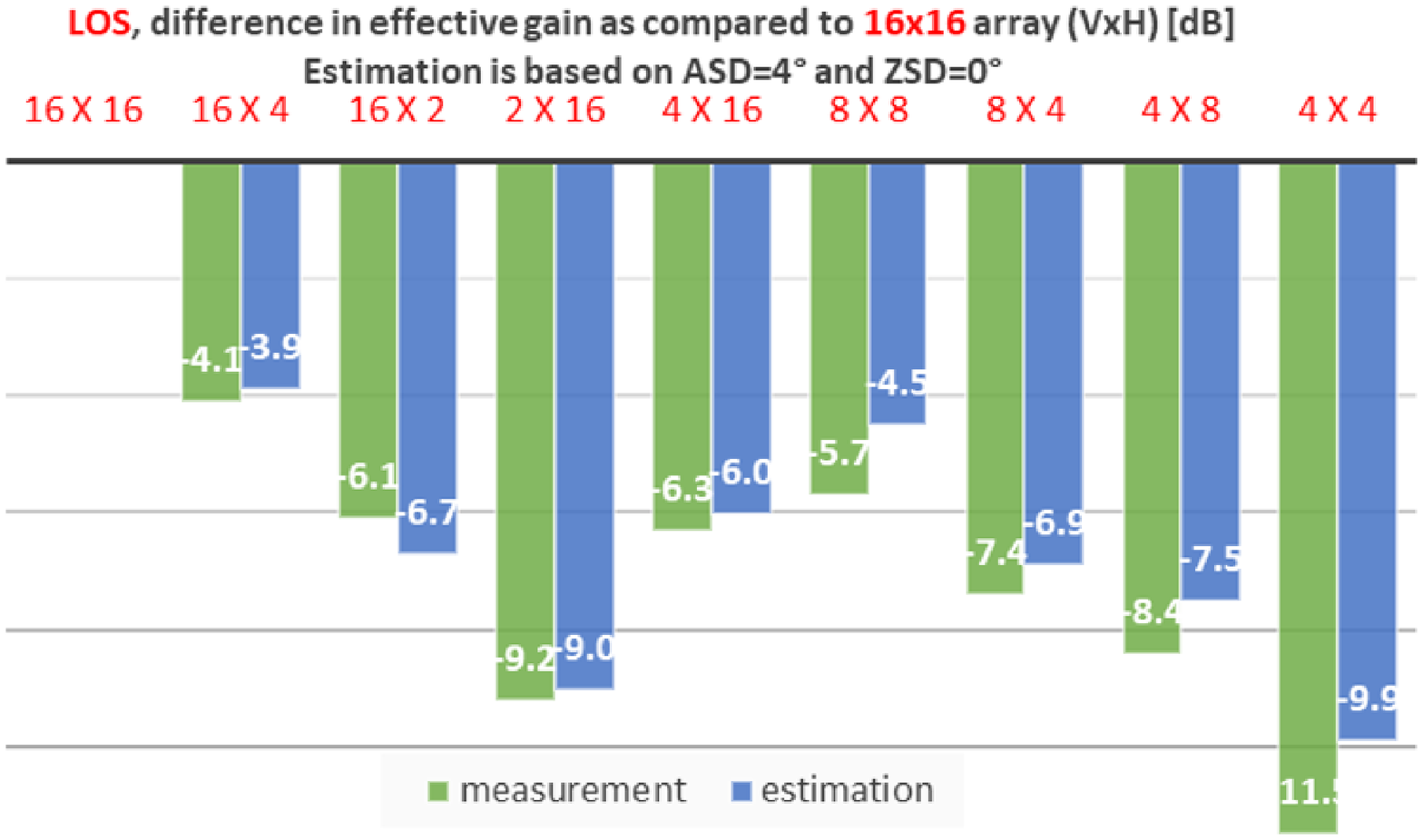}\\
		\includegraphics[width=0.95\figwidth]{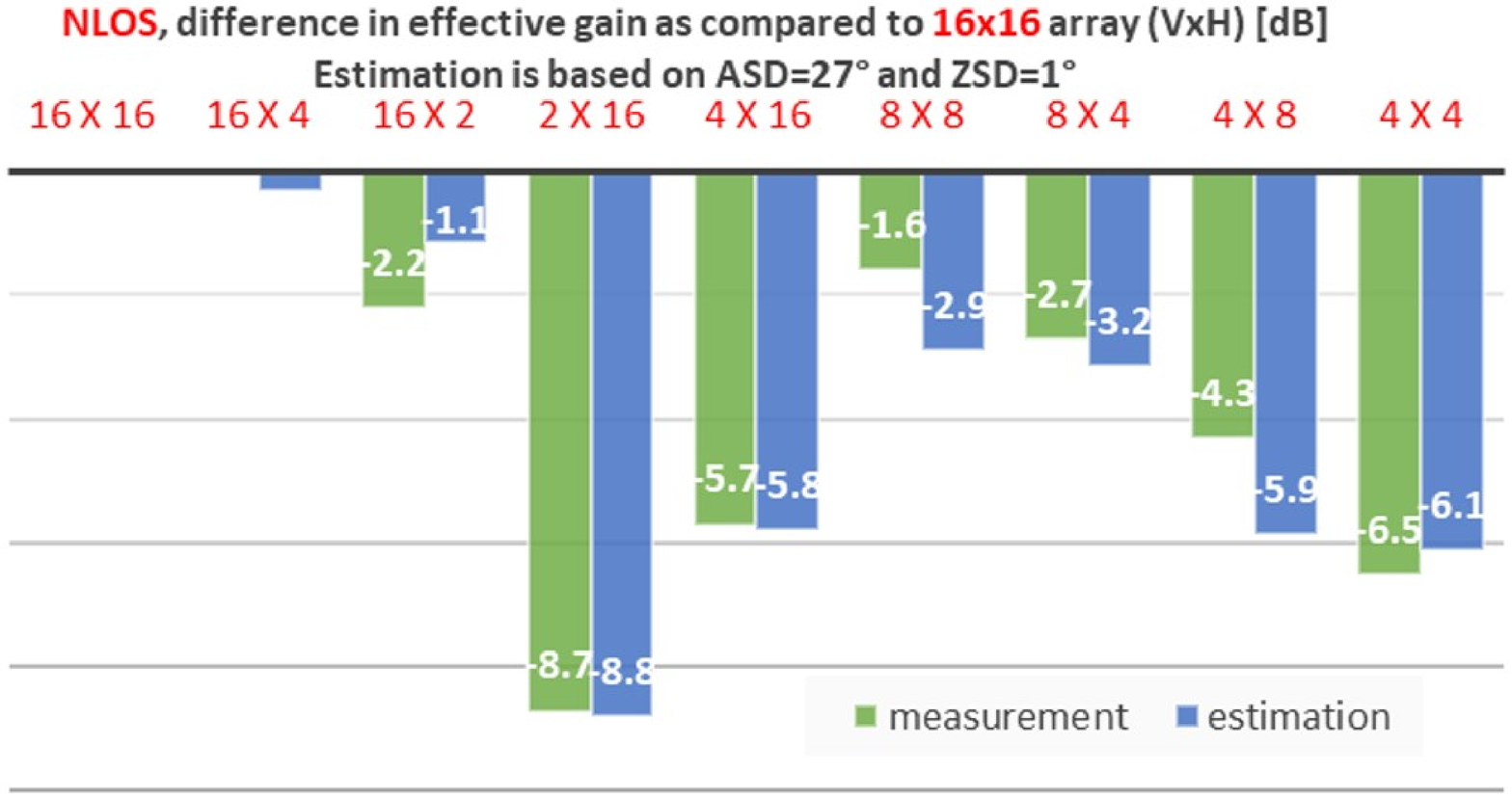}
	\caption{Lab measurement results and estimated effective beamforming gains for LOS (upper) and NLOS (lower).}
	\label{fig:lab_meas_results}
\end{figure}

Lab measurements, as shown in Fig.~\ref{fig:lab_meas_setup}, were carried out for both LOS and NLOS scenarios. For LOS, two rows of reflective panels are used to create multipath-rich environment with larger angular spread in azimuth to verify the gain of optimal antenna arrays. For NLOS measurements,  a metal rack and additional panels are used to increase angular spread. The measured relative gain, using the full $16{\times}16$ array as baseline, as well as the estimated gain based on estimated angular spreads  using the methods presented in Sec.~\ref{sec:spread_estimation} (rounded to integer value) are shown Fig.~\ref{fig:lab_meas_results}.

The results have verified the effective antenna gain for different antenna array geometry with different number of antenna elements for LOS and NLOS scenarios. For example, in LOS, the $16{\times}2$ sub-array has similar gain as the $8{\times}8$ by using 2 times less antenna elements.  In NLOS, the effective antenna gain of $16{\times}2$ array is only 2.2 dB worse than $16{\times}16$, whereas the effective gain of $2{\times}16$ array is $8.7$ dB worse, clearly demonstrated the need of array optimization. Furthermore, these measurement results match our estimated gain (based on estimated angular spread) with high accuracy. These examples clearly validate our analysis on  antenna array optimization and angular spread estimation.      

\section{Potential Applications}\label{sec:application}
 
 We present here a few potential applications where optimizing array geometry can be applied to improve system performance.

\subsection{Deployment Specific Array Optimization}

	\begin{figure} 
	\centering
		\includegraphics[width=0.95\figwidth]{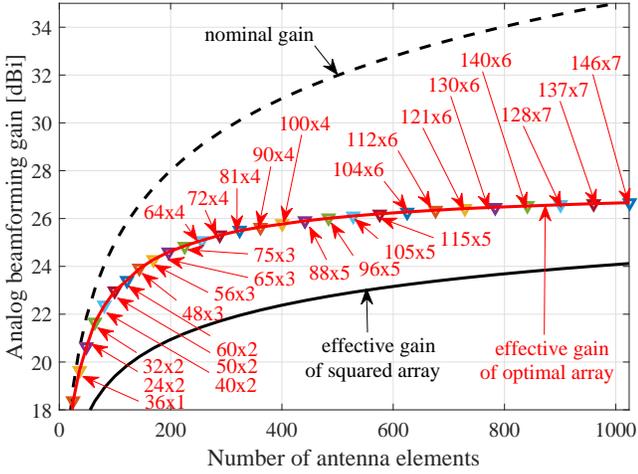}
	\caption{Example of optimal array geometry and the effective gain as function of array size. The ASD and ZSD are according to specifications in 3GPP UMi street canyon NLOS channel~\cite{TWC_5}.}
	\label{fig:opt_array_UMi}
\end{figure}

In environments where azimuth angular spread is much larger than elevation angular spread, which is the case for deployment scenarios covered by 3GPP channel models, a tall array with the same number of elements (e.g., 16$\times$4) may improve the signal strength by a few dB as compared to the square array (i.e., 8$\times$8), thus leading to better performance.  
Pre-design of arrays in different geometry can be targeted for each typical deployment scenarios, such as urban macro sites, urban micro small cells, suburban FWA, and indoor office. For each typical deployment scenario, one may design the array geometry based on the mean value of angular spread in such cases and exploit the fact that the spreads in azimuth and in elevation are not the same. Such design strategy would provide similar gain on SNR over the square array for majority of the users, as verified by our system level simulations. 

In Fig.~\ref{fig:opt_array_UMi} we compare the effective analog beamforming gain of the optimal array to the gain of traditional squared arrays in 3GPP UMi street canyon NLOS deployment scenarios.
Optimal array geometry as labeled in the figure are obtained according to Theorem~\ref{theorem:gain} and the corresponding effective beamforming gain is obtained using \eqref{eqn:effective_gain}. 
For same number of antenna elements,  5 dBi each, the optimal array design can improve the effective beamforming gain (thus the signal strength) by 2 to 3 dB over squared arrays. Configuration for other radio propagation environments with different angular spreads or other values of element gain can be obtained in a similar way straightforwardly.
Since the angular spreads at UE are much larger than those at BS, as shown in Fig.~\ref{fig:3GPP_spread}, using large antenna arrays at UE is inefficient in providing beamforming gain. 



\subsection{Optimizing Array Geometry under EIRP Constraint}

\begin{figure} 
	\centering
		\includegraphics[width=0.95\figwidth]{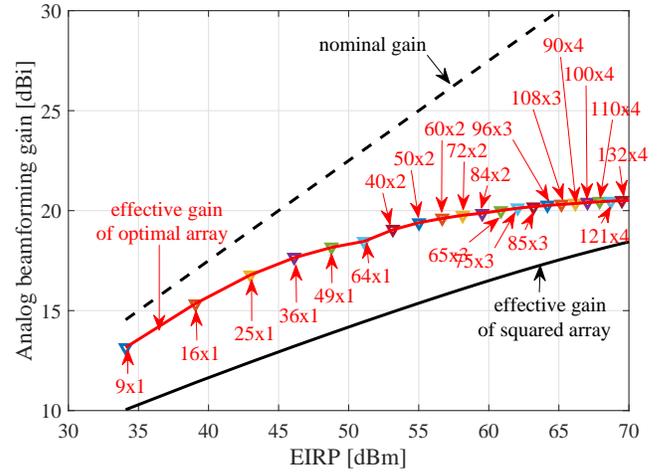} 
	\caption{Example of optimal analog beamforming gain and array geometry as a function of EIRP limit for 3GPP indoor LOS channel~\cite{TWC_5}. 
	}
	\label{fig:opt_array_EIRP}
	\end{figure}

For devices with strict  equivalent isotropic radiated power (EIRP) limit, such as indoor AP/CPE, the maximum allowable number of antenna elements $N$ can be determined from the EIRP limit as:  
\[N\leq10^{(\text{EIRP}- P_t-G_e)/20},\]
where EIRP is in dBm, $P_t$ is the per-element transmit power in dBm and $G_e$ is the per-element gain in dBi. 
For example, with per-element directional gain of 5 dBi and per-element transmit power of 10 dBm, a maximum of 25 elements 
is allowed for indoor mobile stations subject to the peak 43 dBm EIRP limit imposed in the United States~\cite{FCC2016}. At a higher peak EIRP limit of 55 dBm for indoor modems, up to 100 such antenna elements can be used. 
 In Fig.~\ref{fig:opt_array_EIRP} we plot the nominal gain, the effective gain of squared arrays, and the effective gain of optimal arrays with the same number of elements, as a function of EIRP limit, where the optimal configuration of antenna array, obtained by applying Theorem~\ref{theorem:gain}, is as indicated in the figure.  
 Compared to squared arrays with the same EIRP limit, 3 to 4 dB improvement of effective beamforming gain (thus signal strength) can be achieved by array geometry optimization for 3GPP indoor LOS scenarios~\cite{TWC_5}. 
 Configurations for other radio propagation environments with different angular spreads or other values of element gain and element power can be obtained straightforwardly following the same method.

On the other hand, the improved effective gain from array geometry optimization can also be leveraged to maintain the same link budget (thus throughput) but with fewer antenna elements as compared to conventional square arrays. For example, as shown in Fig.~\ref{fig:opt_array_EIRP}, a $5\times 5$ squared array with 43 dBm EIRP (including 24 dBm Tx power) would have effective gain of 13 dBi, whereas a $16\times 1$ array would have 22 dBm Tx power but with effective gain of 15 dBi. Thus, using the $16\times 1$ array would maintain the same link signal strength as the $5\times 5$ squared array but with 2 dB less Tx power and 36\% reduction in antenna elements, which translates to a combined 4 dB reduction of EIRP. Such reduction will not only leads to lower power consumption and reduced hardware cost, but also lower EMF radiation, which could help 5G system to meet performance expectations under RF EMF compliance limits~\cite{EMF_2019}.

\subsection{Array Optimization for FWA Cell Capacity Enhencement}

High path loss and large signal bandwidth (in the order of 1000 MHz) at mmWave bands lead to low to medium SNR for users in NLOS or at long distance. Since the throughput is close to linear of SNR level in noise limited systems, a modest gain in signal strength could lead to substantial gain in throughput, especially for cell edge users.

\begin{figure} 
	\centering
		\includegraphics[width=0.98\figwidth]{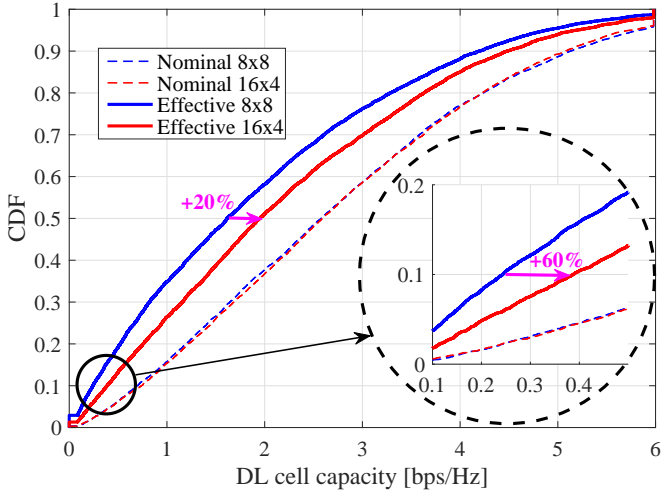} 
	\caption{CDF of DL cell capacity (bps/Hz) for 5G FWA at 28 GHz in a suburban deployment scenario~\cite{5GWF-2019}, where optimized array geometry of $16{\times}4$ is compared to the default $8{\times}8$ squared array. }
	\label{fig:opt_array_FWA}
\end{figure}

In Fig.~\ref{fig:opt_array_FWA} we plot the CDFs of the DL cell capacity (bps/Hz) for 5G FWA at 28 GHz in a suburban residential deployment scenario~\cite{5GWF-2019} where antenna arrays of 64 elements are used at lamppost-mounted access points. Detailed simulation setup can be found in~\cite{5GWF-2019}. With 800 MHz bandwidth and 285 m inter-site distance along the same street, the system is essentially noise limited for most of the Customer Premise Equipment (CPE). The optimized array of $16{\times}4$ achieves about 2 dB gain in median DL SINR as compared to the default $8{\times}8$ squared array. We map the DL SINR to DL cell capacity using the 3GPP configuration~\cite{TWC_5b}, and the plot the CDFs of cell capacity in Fig.~\ref{fig:opt_array_FWA}. As compared to the default squared array, the optimized array provides a 20\%  increase of  cell capacity  at median and 60\% increase at 10th percentile (i.e., cell edge).

%

\section{Conclusions and Discussions}\label{sec:conclusion}

  In this paper we address the link budget challenge of high speed wireless access at high bands by focusing on the effective beamforming gain of antenna arrays under channel angular spread. We have presented closed form solution to match the antenna beam pattern with channel angular spread, which can be very useful in designing deployment specific antenna arrays for typical scenarios based on long-term historical data to improve link budget. We have also developed a method to estimate channel angular spread based on as few as three power measurements, which facilitate dynamic directional beam configuration in a per-transmission basis. This opens the door of a new operation regime for analog beamforming at high frequencies.

Although we made a few assumptions regarding the angular-power distribution to make analysis tractable, the feasibility and projected gains of our methods have been confirmed with impressive accuracy by our 3GPP compliant system level simulations using 3D channel models and by our lab measurement using a 16$\times$16 phased array at 28 GHz.  Furthermore, our proposed use cases for deployment-specific array geometry optimization only require the average value of RMS angular spread, which can be estimated based on historical data for each deployment scenarios.

Since the key ingredients of our solution is to match the beam pattern with channel angular spread, the proposed geometry optimization and angular spread estimation methods also apply to other array types and  beamforming methods,  despite that our description focused exclusively on beamforming over uniform planar array. For such applications, it is the RMS beamwidths in azimuth and in elevation that should be used in analysis rather than the dimension of arrays. 	
The capability of real-time link-specific optimal beam pattern determination developed here is  especially interesting for advanced beamforming techniques of phased arrays~\cite{PhasedArray2018} and  novel antenna technologies using metasurfaces~\cite{MetaSurface2016}.

Extension to panel-based hybrid beamforming is straightforward. Assuming there are in total $N$ antenna elements evenly allocated to $M$ sub-panels, each supported by one dedicated RF chain. Each sub-panel has $N/M$ elements  arranged in rectangular/square shape to form a uniform planar array, where the optimal array geometry $(K_1, K_2)$ can be optimized as in Sec.~\ref{sec:optimization} to maximize the effective analog beamforming gain $G(K_1,K_2)$ for each sub-panel. Assuming perfect CSI is available for digital beamforming when combining M panels via maximum ratio combining/transmission, the effective beamforming gain of the $N$-element $M$-subpanel hybrid beamforming is therefore $MG(K_1,K_2)$.

\section*{Acknowledgment}
The authors would like to thank Dmitry Chizhik for helpful discussions on channel angular spread, and Jakub Bartz for help during all the measurements in the laboratory. 
\appendices

\section{Proof of Optimal Array Geometry}\label{app:Proof}

Assume each antenna element has nominal beamwidth $B_{ve}$ in elevation and $B_{he}$ in azimuth, which could be measured from anechoic chamber. They can also be derived from its nominal gain by assuming identical beamwidth in elevation and in azimuth, i.e., 
	\[B_{ve} = B_{he}=\sqrt{{2}/{G_e}},\]
	where $G_e$ is the gain of the antenna element and the last step is from \eqref{eqn:3}.
	
In free space or anechoic chamber where there is no angular spread, the analog beams formed by an antenna array of size $(K_1, K_2)$ shall preserve its ideal RMS beamwidths $B_{v0}$ and  $B_{h0}$, 
\begin{align}
B_{v0}= \frac{B_{ve}}{K_1},  \   B_{h0} =\frac{B_{he}}{K_2}. \label{eqn:6}
\end{align}
Given angular spread $\sigma_v$ and $\sigma_h$, the effective analog beamforming gain can be determined by substituting \eqref{eqn:6} and \eqref{eqn:4} into \eqref{eqn:3},  described as follows
\begin{align}
  G&(K_1,K_2, B_{ve},B_{he},\sigma_v,\sigma_h ) = \frac{2}{B_v B_h} \nonumber\\
& =\frac{2}{\sqrt{(\frac{B_{ve}}{K_1} )^2 {+} \sigma_v^2 } \sqrt{ (\frac{B_{he}}{K_2})^2 {+} \sigma_h^2}}, \label{eqn:effective_gain}\\
& =\frac{2}{\sqrt{ \frac{B_{ve}^2 B_{he}^2}{K_1^2K_2^2} +\sigma_v^2\sigma_h^2  +  \sigma_h^2\frac{B_{ve}^2}{K_1^2} +\sigma_v^2 \frac{B_{he}^2}{K_2^2} }}. \nonumber
\end{align}

Since $K_1K_2 \leq N$, the effective beamforming gain \eqref{eqn:effective_gain} can be rewritten as
\begin{align}
G & = \frac{2}{\sqrt{ \frac{B_{ve}^2 B_{he}^2}{N^2} +\sigma_v^2\sigma_h^2  +  \sigma_h^2\frac{B_{ve}^2}{K_1^2} +\sigma_v^2 \frac{B_{he}^2}{K_2^2} }} \label{eqn:A2}\\
& \leq \frac{2}{\sqrt{ \frac{B_{ve}^2 B_{he}^2}{N^2} +\sigma_v^2\sigma_h^2  +  2\sigma_h\sigma_v \frac{B_{ve}B_{he}  }{N}   }}   \label{eqn:A3}\\
& = \frac{2}{\sigma_h\sigma_v + \frac{B_{ve}B_{he} }{N}}, \label{eqn:A4}
\end{align}
where \eqref{eqn:A2} is by substitution of $K_1K_2{=}N$, and \eqref{eqn:A3} is from the inequality of arithmetic and geometric means (i.e., the AM-GM inequality), with equality hold, thus achieving the maximal effective gain \eqref{eqn:A4}, if and only if 
\begin{align}
 \frac{  K_1 }{K_2 } =   \frac{\sigma_h B_{ve} }{\sigma_vB_{he} } = \frac{\sigma_h/B_{he}}{\sigma_v/B_{ve}}. \label{eqn:A5}
\end{align}
Combine \eqref{eqn:A5} with constraint $K_1K_2{=}N$ leads to the solution presented in \eqref{eqn:opt_geometry}.


\end{document}